\documentclass[a4paper]{jpconf}
\usepackage{graphicx}
\begin{document}
\title{Strong Reduction of the Field-Dependent Microwave Surface Resistance in YBCO with BaZrO$_3$ Inclusions}

\author{N Pompeo$^1$, V Galluzzi$^2$, A Augieri$^2$, G Celentano$^2$,  T Petrisor$^3$, R Rogai$^1$ and E Silva$^1$}

\address{$^1$ Dipartimento di Fisica ``E. Amaldi'' and Unit\`a CNISM,
Universit\`a Roma Tre, Via della Vasca Navale 84, I-00146 Roma,
Italy}
\address{$^2$ ENEA-Frascati, Via Enrico Fermi 45, 00044 Frascati, Roma, Italy}
\address{$^3$ Department of Physics, Technical
University of Cluj, 400020 Cluj-Napoca (Romania)}

\ead{silva@fis.uniroma3.it}

\begin{abstract}
We present measurements of the magnetic field dependent microwave surface resistance in laser-ablated YBa$_2$Cu$_3$O$_{7-\delta}$ films on SrTiO$_3$ substrates. BaZrO$_3$ crystallites were included in the films using composite targets containing BaZrO$_3$ inclusions with mean grain size smaller than 1 $\mu$m. X-ray diffraction showed single epitaxial relationship between BaZrO$_3$ and YBa$_2$Cu$_3$O$_{7-\delta}$. The effective surface resistance was measured at 47.7 GHz for 60$< T <$90 K and 0$< \mu_0H <$0.8 T.
The magnetic field had a very different effect on pristine YBa$_2$Cu$_3$O$_{7-\delta}$ and YBa$_2$Cu$_3$O$_{7-\delta}$/BaZrO$_3$, while for $\mu_0H=$0 only a reduction of $T_c$ in the YBa$_2$Cu$_3$O$_{7-\delta}$/BaZrO$_3$ film was observed, consistent with dc measurements. At low enough $T$, in moderate fields YBa$_2$Cu$_3$O$_{7-\delta}$/BaZrO$_3$ exhibited an intrinsic thin film resistance lower than the pure film. The results clearly indicate that BaZrO$_3$ inclusions  determine a strong reduction of the field-dependent surface resistance. From the analysis of the data in the framework of simple models for the microwave surface impedance in the mixed state we argue that BaZrO$_3$ inclusions determine very steep pinning potentials.
\end{abstract}

\section{Introduction}
The search for efficient pinning centers in high-$T_c$ superconductors was started immediately after it was recognized that in a very wide portion of the $H-T$ phase diagram flux lines were able to move, and very difficult to pin. Up to recent years, the strongest pinning centers were identified with columnar defects \cite{civalePRL91}, produced by heavy-ion irradiation. The very efficiency of such defects in pinning vortex lines is counterbalanced by the complexity of the technique, which is impractical in view of large-scale production of high-$T_c$ superconductors for applications. Recently, it has been shown \cite{macmanusNATMAT04, kangSCI06} that micrometer and nanometer sized BaZrO$_3$ (BZO) crystallites could dramatically improve the dc properties, such as critical currents and irreversibility line of YBa$_2$Cu$_3$O$_{7-\delta}$ (YBCO) films, and irreversibility fields above 10 T at 77 K were reached \cite{PeurlaPRB07}. Such features could well be of interest in the field of the signal applications at high frequencies (microwaves). While it is unlikely that radiofrequency or microwave devices should operate in fields of several Tesla, nevertheless a potential capability to reduce the losses in moderate magnetic fields (tenths of tesla) calls for further investigation. Moreover, it has been shown \cite{ghigoSUST05} that strong pinning centers such as columnar defects, when suitably distributed, may significantly extend the range of linear response (power handling) by acting as pinning centers for fluxons nucleated by the rf field. However, the efficiency of BZO in this respect remains to be investigated.\\
The dynamics of flux lines at microwave frequencies differs from dc properties because different features of the pinning potential are probed. Microwaves induce very small oscillations of the vortex around its equilibrium position (fractions of nm \cite{TomaschPRB88}). At the high edge of the microwave spectrum (40-100 GHz) it is fair to state that the dynamics of single vortices is probed. In particular, the real part of the vortex resistivity (dissipation) is mainly given by viscous motion, while the imaginary part (elastic vortex response) is mainly given by the steepness of the pinning potential. Below the irreversibility line, where flux creep can be neglected, one can invoke the well-known Gittleman-Rosenblum (GR) model \cite{gr}:
\begin{equation}
    \Delta\rho_1+\mathrm{i}\Delta\rho_2=\rho_{ff}\frac{1+ \mathrm{i} \frac{\nu_p}{\nu}}{1+\left(\frac{\nu_p}{\nu}\right)^2}
    \label{eqGR}
\end{equation}
%
%
%
where $\rho_{ff}$ is the flux-flow resistivity, $\nu_p\propto k_p$ is the depinning frequency and $k_p$ is the pinning constant which measures the steepness of the pinning potential. Within the applicability of this model, microwaves probe the steepness of the pinning potential and the viscous motion only.
\section{Film preparation and characterization}
Films were prepared by Pulsed Laser Deposition technique using a $\lambda$=308 nm XeCl excimer laser with a repetition rate of 10 Hz. The laser energy density was about 2.5 J/cm$^{2}$ and the substrate-target distance was 4.8 cm. Targets for laser ablation were obtained by mixing submicrometric YBCO and BZO powder and then sintering, resulting in a composite YBCO/BZO target with nominal BZO content of 7 molar percent. Pure YBCO films were obtained using commercially available high density laser ablation targets. Deposition temperature and oxygen partial pressure for both YBCO and YBCO/BZO were 850 $^{\rm o}$C and 300 mTorr. The thickness of the as deposited samples was about 130 nm.

\begin{figure}[t]
\begin{center}
\includegraphics[width=16.5cm]{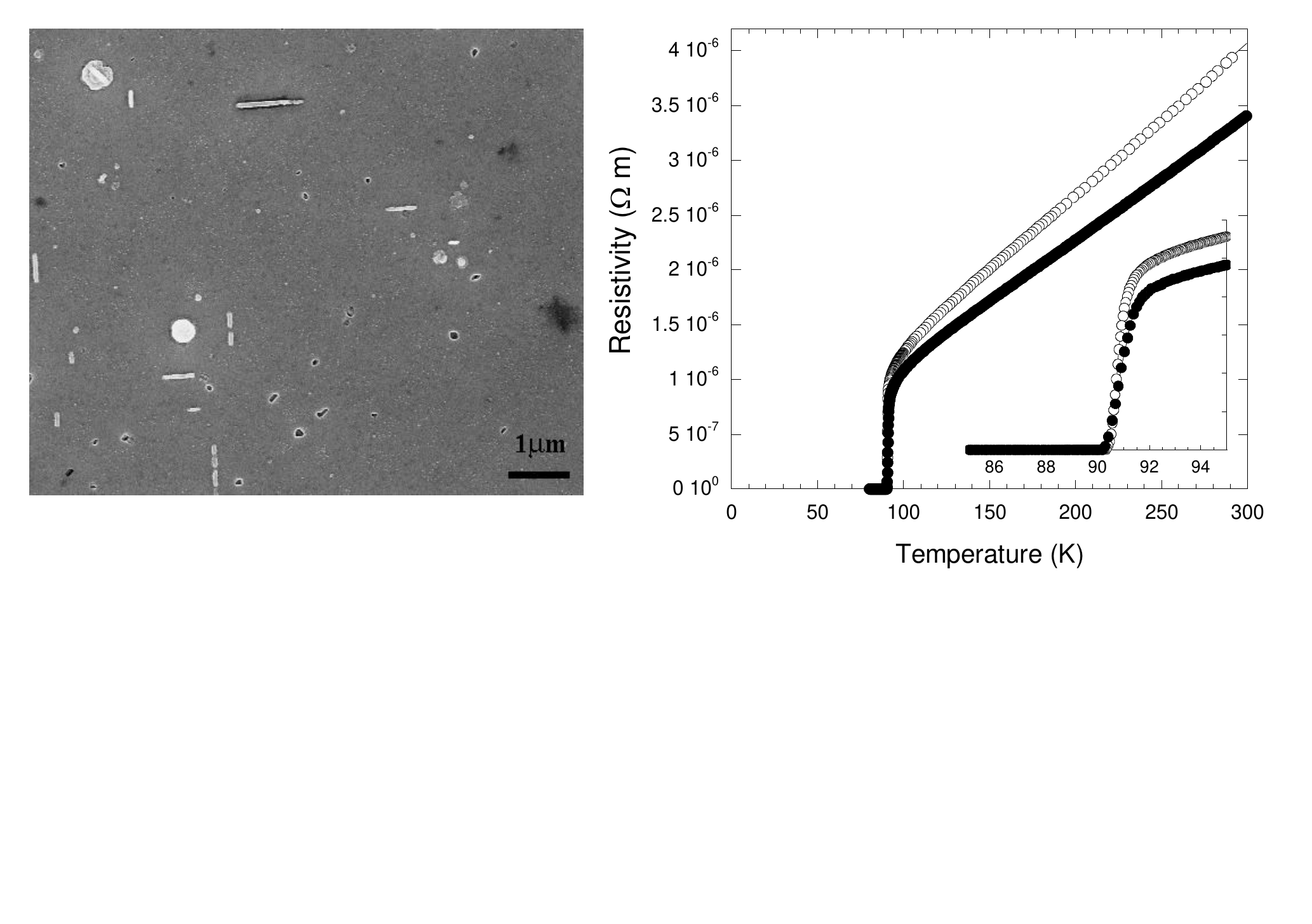}
\end{center}
\vspace{-5cm}
\caption{\label{SEMrho} {\it Left panel}: SEM analysis on YBCO/BZO sample, showing a very smooth and flat surface. {\it Right panel}: DC resistivity $\rho$ in pure YBCO (~\opencircle) and in YBCO/BZO (~\fullcircle): the slight reduction of $\rho$ in YBCO/BZO with respect to pure YBCO is probably of intrinsic nature.}
\end{figure}
X-ray spectra performed by a Rigaku-Geigerflex diffractometer with Cu-K$\alpha$ radiation in $\theta-2\theta$ configuration show that YBCO films grew epitaxially on the (001) STO substrate, revealing a (00\textit{l})-oriented YBCO film with (005) rocking curve FWHM of about 0.2$^{\rm o}$ for pure YBCO samples and 0.16$^{\rm o}$ for YBCO/BZO samples. In the YBCO/BZO film the BZO crystallites are (\textit{h}00) oriented, indicating single epitaxial relationship between BZO and YBCO.
The morphology for all the samples was observed through SEM microscopy. As it can be noticed in left panel of figure \ref{SEMrho}, the typical spiral growth observable on YBCO films is not seen on the YBCO/BZO surface, suggesting a very flat and smooth surface.

Measurements of DC resistivity $\rho$ versus temperature (right panel of figure \ref{SEMrho}) yield, at 300 K, $4.1\times10^{-4}$ cm and $3.4\times10^{-4}$ cm for YBCO and YBCO/BZO, respectively. The DC resistivity in HTS films is affected by different factors such as the film orientation and degree of epitaxy, the intergrain connectivity and intragrain intrinsic properties, such as possible cation disorder in the film structure or any change in oxygen stoichiometry. The reduced value of $\rho$ measured for BZO-added films is here not to be ascribed to a difference in texture quality as it can be deduced from X-ray data, being more likely related to the different intrinsic resistivity values.
A very high zero-resistance critical temperature $T_{C0}$ value of 89.9 K is obtained in the YBCO/BZO film, but still reduced if compared (right panel of figure \ref{SEMrho}, inset) to pure YBCO in which $T_{C0} = 90.2$ K as already reported also in previous works \cite{galluzziIEEE07}.\\
\begin{figure}
\begin{center}
\includegraphics[width=12cm]{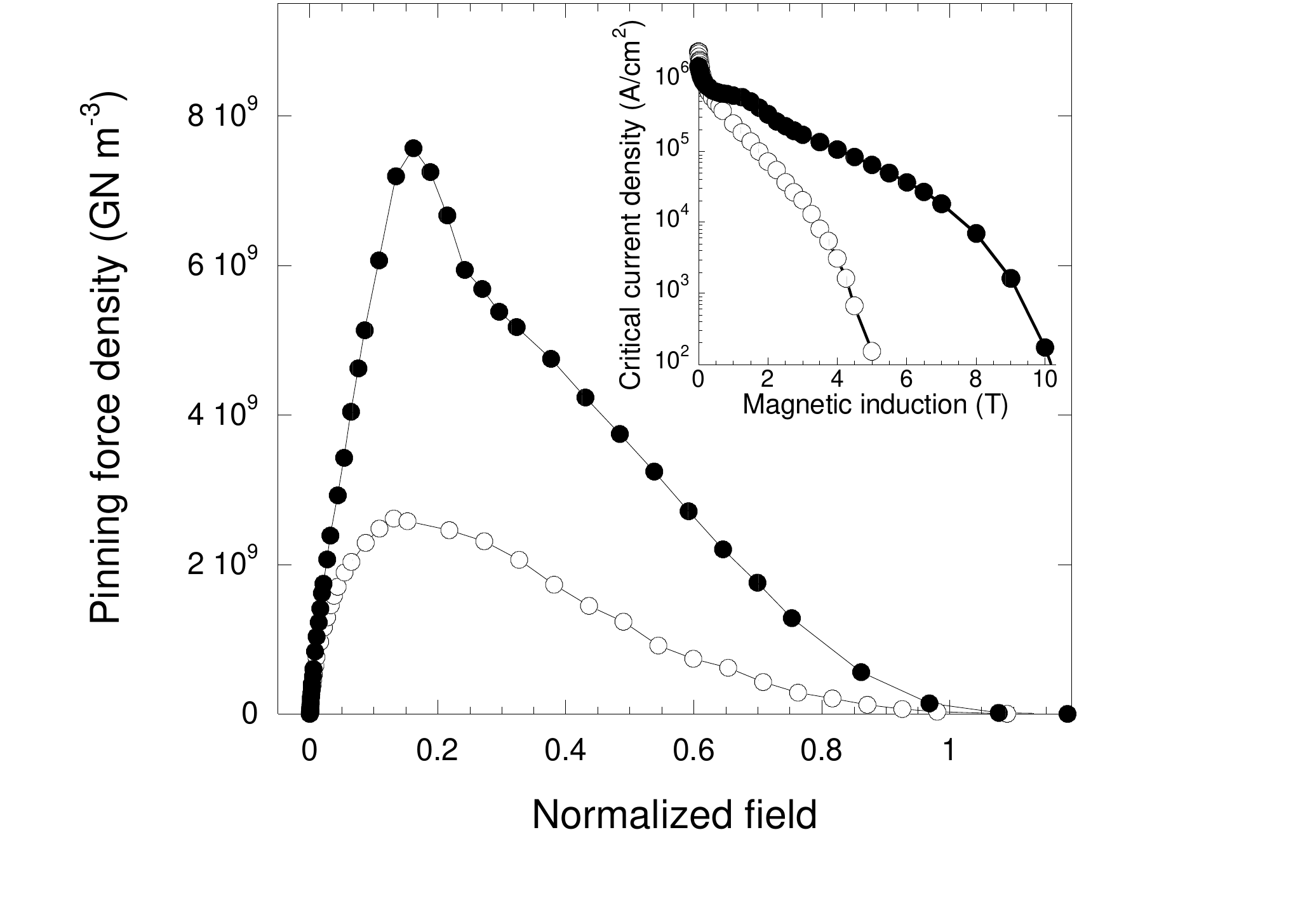}
\end{center}
\vspace{-1cm}
\caption{\label{Fp} Field dependence of the pinning force density (main  panel) and critical current density (inset) in pure YBCO (~\opencircle) and in YBCO/BZO (~\fullcircle), evidencing a net improvement of the pinning properties in the latter sample.}
\end{figure}
\\
Both samples were mounted in a He gas flow cryostat provided with a superconducting magnet and in-field \textit{I-V} curves were recorded at 77 K using the four-point method. The $J_{c}$ values were evaluated using the voltage criterion of 1 $\mu$V cm$^{-1}$. Although BZO added films exhibited a self field $J_{c}=1.58$ MA/cm$^{2}$, which is only slightly lower than the value 2.50 MA/cm$^{2}$ obtained for the pure YBCO sample, a remarkable difference in the in-field performances is evidenced in the inset of figure \ref{Fp} where magnetic field dependencies of the $J_{c}$ values are plotted.\\
BZO film shows a better magnetic field retention especially in the low field regime where a plateau, not present in the pure YBCO, can be noted up to 1 T. Evaluating the irreversibility field $H_{irr}$ as the field where the pinning force density $F_{p}=J_{c}B$ decreases to the 1$\%$ of its maximum, the improvement of the in field performances can be quantified. The resulting $H_{irr}$ values are 9.3 T and 4.6 T for BZO added YBCO and pure YBCO films respectively.
From the analysis of pinning force densities (figure \ref{Fp}) as calculated from transport data some considerations can be done. A larger maximum pinning force density, 7.6 GN/m$^{3}$, is exhibited by BZO film compared with 2.6 GN/m$^{3}$ exhibited by pure YBCO film, confirming that BZO addition improves the pinning efficiency. The difference between the shapes of $F_{p}(H/H_{irr})$ curves in YBCO/BZO and pure YBCO films gives an indication that the improvement can be ascribed to a new pinning mechanism introduced in BZO samples.

\section{Microwave measurements}
We measured the field-dependent microwave response at 47.7 GHz of a pure YBCO film and of an YBCO/BZO (sample 7\%). Measurements were taken by the dielectric resonator technique described in \cite{pompeoJSUP07}. The superconducting film occupies an end-wall of a cylindrical enclosure loaded with a sapphire rod. The quality factor $Q$ of the structure measures the overall losses (the higher the $Q$, the lower the losses), while field-induced changes in $Q$ and in the resonant frequency $f_0$ measure the field-induced variation of the losses and of the reactance due to the superconductor only. The response is affected by substrate resonances given by the strongly varying permittivity of the STO substrate \cite{kleinJAP90, silvaSUST96}.  In order to remove the spurious response we added a thin dielectric spacer to tune the substrate impedance as described in \cite{pompeoPREP07}. Thus, all the measurements here presented are free from such spurious contributions. In that case the thin film approximation held, so that $Q$ and $f_0$ yield an effective surface impedance $\Delta Z_s'\propto\Delta\rho_1+\mathrm{i}\Delta\rho_2$.\\
\begin{figure}[t]
\begin{center}
\includegraphics[width=16.5cm]{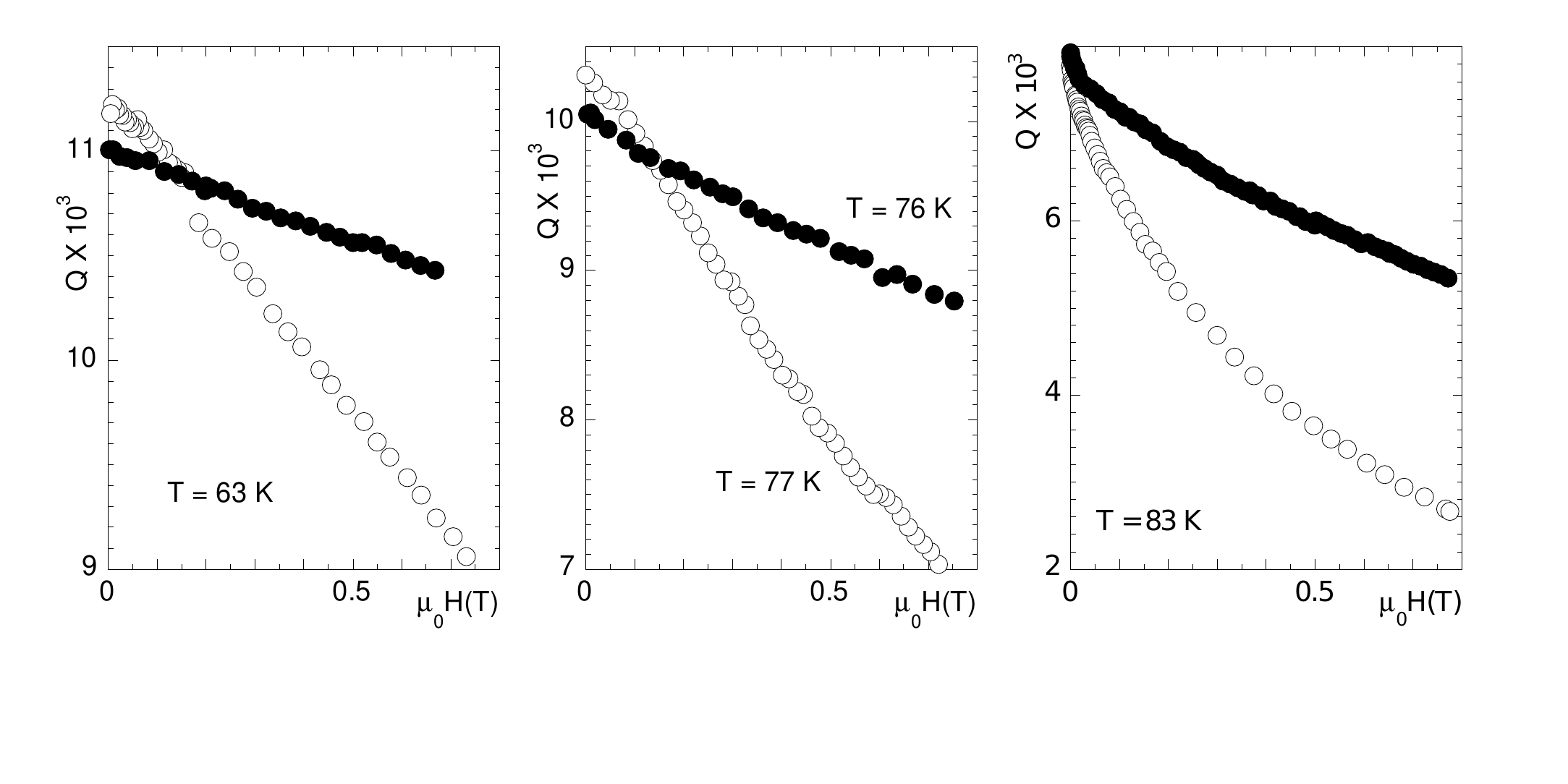}
\end{center}
\vspace{-2cm}
\caption{\label{figQ}Field dependence of the losses ($Q$ factor) in pure YBCO (~\opencircle) and in YBCO/BZO (~\fullcircle) at nearly same temperatures. With increasing field the losses in YBCO/BZO become smaller (larger $Q$) than in pure YBCO.}
\end{figure}
A dc magnetic field applied along the c-axis of the films was swept from 0 up to $\mu_0 H = $0.8 T at different temperatures in the range 60 - 90 K.\\
In figure \ref{figQ} we present the main experimental findings of this Section. The figure reports the $Q$ factor of the resonator incorporating the superconducting film as a function of the magnetic field, at some representative temperatures. It is apparent that, while the zero-field losses are slightly smaller in pure YBCO (larger $Q$), the field-induced drop of $Q$ is much stronger: at fields as small as 0.2 T (at 63 K) the YBCO/BZO exhibits lower losses. Moreover, it is clear that the decrease of $Q$ is much faster in pure YBCO than in YBCO/BZO. This finding demonstrates that the introduction of BZO particles has substantially increased the flux pinning even at our high operating frequency. This is an additional feature with respect to the known strong pinning in dc, where motion of flux lines over large distances (or a steady motion in the flux-flow regime) is probed. At microwave frequencies low dissipation arises from small vortex velocity (i.e. small mean vortex displacement in the harmonic regime). Figure \ref{figQ} indicates that BZO particles are very effective in the reduction of the mean vortex displacement, as can arise from very steep pinning wells. This is an additional piece of information with respect to the known efficiency of BZO as pinning centers in dc, which instead implies deep pinning wells.\\

\begin{figure}
\begin{center}
\includegraphics[width=16.5cm]{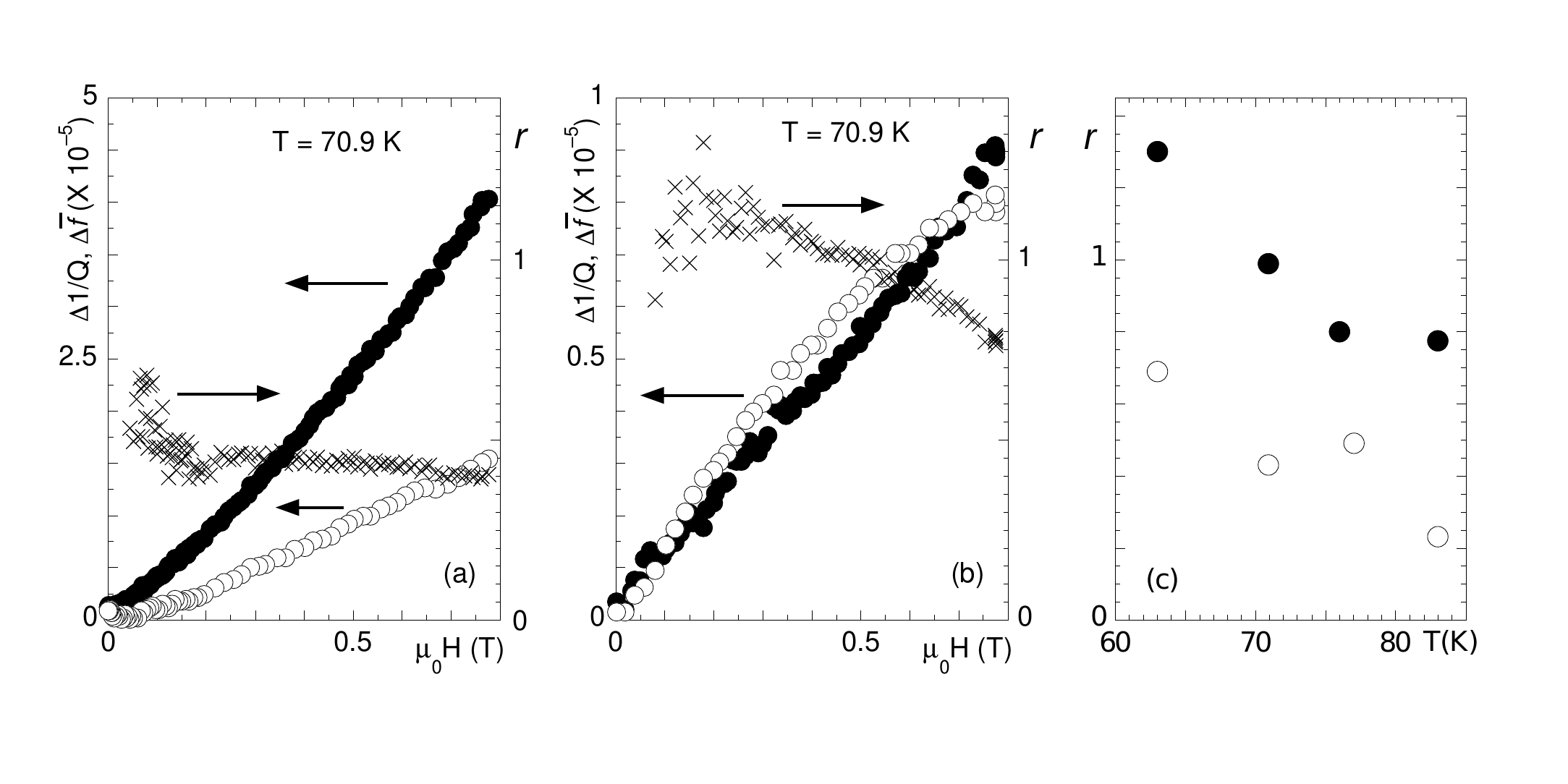}
\end{center}
\vspace{-1.5cm}
\caption{\label{fig_Qfr}Field dependence of $\Delta \frac{1}{Q}$ (losses, \fullcircle), of the reactance ($\Delta\bar f$, \opencircle) and of the $r$ parameter ($\times$) in (a) pure YBCO and (b) YBCO/BZO at $T=$70.9 K. Note the very different vertical scales. (c): temperature dependence of $r$ at $\mu_0H=$ 0.5 T in pure YBCO (\opencircle) and in YBCO/BZO (\fullcircle).}
\end{figure}

In order to further evaluate the pinning properties of BZO, we discuss the behaviour of the well known $r$ parameter \cite{halbritter}. This important quantity is defined as the ratio of the field-variations of the reactance $\Delta X_s'(H)$ and of the resistance $\Delta R_s'(H)$, and it can be directly obtained from the experimental data for $Q(H)$ and $f_0(H)$as follows:
\begin{equation}
    r=\frac{\Delta X_s'(H)}{\Delta R_s'(H)}=\frac{2\frac{f_0(0)-f_0(H)}{f_0}}{ \frac{1}{Q(H)}- \frac{1}{Q(0)}}=\frac{\nu_p}{\nu}
    \label{eq_r}
\end{equation}
Here, the last equality is valid in the GR model, see \ref{eqGR}. We stress that the experimental determination of $r$ is free from any calibration of the resonator.\\
Figures \ref{fig_Qfr}a and \ref{fig_Qfr}b report typical field-induced variations of resistive, $\Delta \frac{1}{Q}=\frac{1}{Q(H)}- \frac{1}{Q(0)}$, and reactive, $\Delta \bar f=2\frac{f_0(0)-f_0(H)}{f_0}$, contributions, and the corresponding $r$ parameter at one intermediate temperature in YBCO and YBCO/BZO. It is seen that in YBCO/BZO the field induced losses are smaller by the large factor $\sim 3-4$ (note the different vertical scales), while the reactive response is smaller by a similar factor $\sim$2.
The $r$-parameter has a slightly more pronounced field dependence in YBCO/BZO than in pure YBCO, but it always attains significantly larger values in YBCO/BZO. It should be mentioned that our pure YBCO was already strongly pinned, as demonstrated by its large $r$. It is noteworthy that BZO further increases the high-frequency pinning efficiency. We also stress that in columnar-irradiated YBCO the reduction of the film surface resistance is of order of 15\% \cite{silvaIJMPB00}, much smaller than the large factor here reported.\\
The comparison of the temperature dependence of $r$, evaluated at $\mu_0H=$ 0.5 T, in YBCO and YBCO/BZO is reported in figure \ref{fig_Qfr}c. It is seen that YBCO/BZO exhibits larger $r$ at all temperatures, indicating that the BZO-induced additional pinning is strong enough to survive to thermal excitation. In this sense, and corroborated by the dc results, we argue that BZO particles induce both very deep and very steep pinning wells in YBCO.

\section{Conclusions}
We have successfully grown YBCO films with different concentrations of sub-$\mu$m size BZO inclusions by pulsed laser ablation, by suitably prepared targets. We have shown that BZO crystallites grew with single epitaxial relationship with respect to YBCO. We have measured the microwave response in pure YBCO and in YBCO/BZO at 7\% mol. as a function of a dc magnetic field. We have shown that the addition of BZO particles only slightly increases the film surface resistance in zero field. With the application of a moderate magnetic field, the YBCO/BZO film shows significantly smaller losses than pure YBCO. We reported a reduction by a factor $\sim 3-4$ of the field-induced surface resistance in YBCO/BZO with respect to YBCO, in agreement with the improvement of the pinning force density observed through DC transport measurements. This reduction is most likely ascribed to the steepness of the pinning wells induced by BZO addtions.

\ack{ringraziamo?}

\end{document}